\documentclass[aip,graphicx]{revtex4-1}

\usepackage{physics}

\draft % marks overfull lines with a black rule on the right

\begin{document}

%\preprint{}

\title{From the Harmonic Oscillator to Time-Frequency Analysis of Chirp Signals} %Title of paper



\author{Donald J. Kouri}
    \affiliation{ 
    Department of Physics, University of Houston, Science and Research Building 1, 3507 Cullen Blvd Room 617, Houston, TX 77204-3008, USA
    }
    \affiliation{ 
    Department of Mathematics, University of Houston, Philip G. Hoffman Hall,
Houston, TX 77204-3008, USA
    }
    \affiliation{ 
    Department of Mechanical Engineering, University of Houston, Engineering Bldg, 4726 Calhoun Rd, Houston, TX 77204-4005, USA
    }
\author{Caleb Broodo}
\affiliation{ 
    Department of Electrical and Computer Engineering, University of Houston, Engineering Bldg 1
4726 Calhoun Rd, Houston, TX 77204-4005, USA
    }  
\author{Bernhard G. Bodmann}
\affiliation{ 
    Department of Mathematics, University of Houston, Philip G. Hoffman Hall,
Houston, TX 77204-3008, USA
    }
\author{Cameron L. Williams}
\affiliation{ 
     Department of Physics, University of Houston, Science and Research Building 1, 3507 Cullen Blvd Room 617, Houston, TX 77204-3008, USA
    }
    \affiliation{ 
    Department of Mathematics, University of Houston, Philip G. Hoffman Hall,
Houston, TX 77204-3008, USA
    }
%\thanks{}
% Collaboration name, if desired (requires use of superscript address option in \documentclass). 
\date{\today}
\begin{abstract}
\begin{abstract}

\end{abstract}
$\textbf{Abstract}$ This paper presents a novel approach to understanding the role of harmonic dynamics and gaining a deeper appreciation for its impact within and outside of quantum mechanics. This includes consequences of harmonic dynamics and the uncertainty principle for anomalous diffusion and for the time-frequency analysis of chirp signals. In this approach, we consider a contact transformation to view a system of canonical variables with coordinate $x$ and momentum $p_x$ in the context of a new system of ``generalized" coordinates and momentum. This new system is first studied in the context of non-relativistic quantum mechanics. The classical analog is then explored by use of the Poisson bracket equation. From this, new implications are demonstrated in classical phenomena. One is for a new model of Anomalous and Normal Diffusion. In another, we introduce the concept of the ``Mixed Fourier Transform" which explores a new Gaussian Fourier Transform kernel in terms of the generalized variables. This has the ultimate objective of ``harmonizing'' chirp signals or producing a harmonic signal from an otherwise non-harmonic chirp.
\end{abstract}


\maketitle %\maketitle must follow title, authors, abstract and \pacs


\section{Introduction}
	Harmonic dynamics is of great interest both for classical and quantum phenomena. It provides prototype model systems that permeate science and engineering, ranging from various sinusoidal motions (harmonic oscillator, pendulum, low displacement motion, and classical oscillations of all sorts, electrodynamics, etc.) %, random processes (normal diffusion, probability theory, statistics, etc.) 
	to quantum mechanics and field theory (Heisenberg's uncertainty principle, molecular interactions, solid state physics, particle physics, etc.). 
	%BGB edit % 
	A system undergoing harmonic dynamics exhibits a characteristic exchange between kinetic and potential energy.
	%One reason for this is the role of kinetic energy in all areas of dynamics. In classical mechanics, kinetic energy is associated with energy due to motion and dimensionally, it involves $sec^2$. This time dependence also occurs in the force-acceleration basis of Newton's law of motion. 
	The most elegant formulations of classical dynamics culminated with Poisson's bracket formulation of the equations of motion, which is built on the dynamics of Newton, Lagrange, and Hamilton\cite{miranda2014introduction}. 
	%The other key ingredient is the energy due to position and in harmonic dynamics, the energy, as a function of position, induces an oscillatory exchange of energy between kinetic and potential forces. 
	In the case of electrodynamics, the origin of periodic behavior arises in a more subtle fashion. Maxwell's equations are relativistically covariant and involve first order time derivatives of the magnetic and electric vector fields. However, upon decoupling $\vec{E}$  and $\vec{B}$, one finds again the appearance of harmonic time evolution due to the time evolution of each being governed by a second time derivative leading to sinusoidal time evolution.
    
	This continues over into quantum mechanics. In the non-relativistic setting, the time dependence deviates from the classical model in the fact that the quantum state, $\ket{\Psi(t)}$, evolves due to a first order time derivative. Ultimately, this reflects the remnants of special relativity and has a connection to the structure of quantum field theory. The fact remains that in quantum theory, harmonic dynamics  %BGB Edit %
	is intimately linked to the Heisenberg Uncertainty Principle (HUP)! Simply put, the fact that $[\hat{x},\hat{p_x}] \neq{0}$ is one of the primary sources of quantum theory's "strangeness". %It is at the heart of the fact that simultaneous exact histories of $x(t), p_x(t)$ are impossible, and that unlike classical dynamical paths being dominated by a least action principle, it states that all possible paths connecting $x(t_i),p_x(t_i)$ and $x(t_p), p_x(t_p)$ contribute, but with a strong phase dependence \cite{ShankarPQM}.
    
    While the HUP fundamentally separates purely classical dynamics from quantum, the surprising fact is that it also is reflected in classical wave dynamics (which for classical waves not arising from electrodynamics, is firmly based on Newtonian equations of motion). Thus, for classically based waves, one knows that such waves cannot be simultaneously well localized in positional wave number or in time and frequency: $\Delta{x}\hbar\Delta{k}\geq{\frac{1}{2}}$, $\Delta{t}\Delta{\omega}\geq{\frac{1}{2}}$.These inequalities are true because of the incommutability of the variables time and frequency or position and momentum (wave-number). The HUP is present with Fourier Transform pairs. This occurs in quantum mechanics and in the time-frequency analysis\cite{book}.   
    
    In recent work, we have demonstrated that a rigorous mathematical derivation of Fourier analysis can be based directly on $\Delta{x}\Delta{k}\geq{\frac{1}{2}}$ quantal (or $\Delta{\omega}\Delta{t}\geq{\frac{1}{2}}$ classical). A bi-product of our analysis is the role of the quantum state that minimizes $\Delta{x}\Delta{k}$: the Gaussian. A fundamental implication of $\Delta{x}\Delta{k}\geq{\frac{1}{2}}$ is that 
    \begin{equation} \label{1}
    \hat{x}\ket{\Psi_{min}} =-i\hat{p_x}\ket{\Psi_{min}}
    \end{equation}
This is an abstract operator relation that states that $\ket{\Psi}$ has the exact form in either the $\ket{x}$- or $\ket{p_x}$- representations. Choosing a representation in quantum mechanics really has to do with looking at the set of observables and deciding which one(s) will be treated as outcomes while the other incompatible observable is only present implicitly. This, in the $\ket{x}$ - representation, $\hat{x}$ is treated classically and $\hat{p_x}$ is treated as an operator:
\begin{equation} \label{2}
\hat{x} =x,\hat{p_x} =-i\frac{d}{dx},			 (\hbar = 1)
\end{equation}
and in the $\ket{p_x}$ - representation, $\hat{p_x}$ is treated classically and therefore $\hat{x}$ is an operator:	
\begin{equation} \label{3}
\hat{x}=i\frac{d}{dx},\hat{p_x}=\hbar k =p_x
\end{equation}
When either of these is used in eq.(\ref{1}), we find
\begin{equation} \label{4}
\Psi_{min}(x) =Ce^{\frac{-x^2}{2}}
\end{equation}
\begin{equation} \label{5}
\Psi_{min}(x) =Ce^{\frac{-p_x^2}{2}} = e^{\frac{-k^2}{2}}
\end{equation}
Even more interesting is the fact that for the harmonic oscillator, the abstract Hamiltonian operator is 
\begin{equation} \label{6}
\hat{H} = \frac{\hat{p_x}^2}{2}+\frac{\hat{x}^2}{2}
\end{equation}
This implies that in the $\ket{x}$ - representation,
\begin{equation} \label{7}
\hat{H} = -\frac{1}{2}\frac{d^2}{dx^2} + \frac{x^2}{2}
\end{equation}
and in the $\ket{k}$ - representation,
\begin{equation} \label{8}
\hat{H} = -\frac{1}{2}\frac{d^2}{dk^2} + \frac{k^2}{2}
\end{equation}
The Hamiltonian is also force invariant in either the $\ket{x}$ - or  $\ket{k}$ - representation. This underlies eqs. (\ref{4})-(\ref{5}) and says
\begin{equation} \label{9}
\ket{\Psi_{0,HO}} \equiv{\ket{\Psi_{min,HO}}}
\end{equation}
In fact, the $\ket{\Psi_{min}}$ state is the state that is most classical in the sense that $\Delta{x}\Delta{k}>0$ is the minimum possible value! We believe that this is a fundamental fact that dictates the ground state for any reasonable system in quantum mechanics. Thus, $\ket{\Psi_{0}}$ is that state which gives the minimum uncertainty for some generalized canonical Cartesian momentum and either $x$-momentum, $p_x$, or a generalized canonically conjugate momentum. This suggests that, beginning with canonical variables $x$, $k=p_x$, one can do a canonical transformation to generalized coordinates $W(x)$, $P_W$, such that for the hypothetical Harmonic Oscillator,

\begin{align} \label{10}
\hat{H}_W &= \frac{\hat{P}_W^2}{2} + \frac{\hat{W}^2}{2} \\ \label{11}
&= -\frac{1}{2}\frac{d^2}{dW^2} + \frac{{W}^2}{2}  \\ \label{12}
&= -\frac{1}{2}\frac{d^2}{dW^2} + \frac{P_W^2}{2} 
\end{align}
the ground state is 
\begin{align} \label{13}
\Psi_{0}(W) &= Ce^{-\frac{W^2}{2}} \\ \label{14}
\widetilde{\Psi}_{0}(P_W) &= Ce^{-\frac{P_W^2}{2}} = Ce^{-\frac{K^2}{2}} 
\end{align}
and
\begin{align} \label{15}
\Psi_{min}(W) &= Ce^{-\frac{W^2}{2}} \\ \label{16}
\widetilde{\Psi}_{min}(P_W) &= Ce^{-\frac{K^2}{2}}
\end{align}
Indeed,
\begin{align} \label{17}
\hat{W}\ket{\Psi_{min}} &= -i\hat{P}_W\ket{\Psi_{min}} \\ \label{18}
\hat{W}\ket{\Psi_{0}} &= -i\hat{P}_W\ket{\Psi_{0}}
\end{align}
Clearly, in the $W$-representation,
\begin{align} \label{19}
\hat{P}_W &= -i\frac{d}{dW}, \\ \label{20}
\hat{W} &= W
\end{align}
and in the $P_W$-representation,
\begin{align} \label{21}
\hat{W} &= -i\frac{d}{dP_W}, \\ \label{22}
\hat{P}_W &= P_W = K
\end{align}
Then it immediately follows that there exists a generalized Fourier transform
\begin{equation} \label{23}
\braket{W}{P_W} = \frac{e^{iWK}}{\sqrt{2\pi}}, \braket{P_W}{W} = \frac{e^{-iWK}}{\sqrt{2\pi}}
\end{equation}
where $WK$ is dimensionless.
We now ask what are possible canonical transformation $W(x)$, $P_W(x,p_x)$ (note, we consider only contact transformation). The most important issue is that at some point, we must quantize the resulting canonically conjugate variables $W$, $P_W$. This is straight forward provided $W$, $P_W$ are Cartesian canonical variables \cite{Dirac1930-DIRTPO}. We take this to require that the domains are
\begin{align} \label{24}
-\infty < W(x) < \infty \\ \label{25}
-\infty < P_W(x,p_x) < \infty 
\end{align}
and that $W(x)$ must be a strictly monotonic function of $x$ (invertible). One transform that ensures this is 
\begin{equation} \label{26}
W(x) = \sum_{j=1}^{J}a_jx^{2j-1}, a_j \geq{0}
\end{equation}
Then 
\begin{equation} \label{27}
\frac{dW}{dx} = W' > 0 \ a.e. 
\end{equation}
If we require $a_1\ne{0}$, then $W$  is invertible. This next step marks the transition to classical mechanics. We generate $P_W(x,p_x)$ by solving the Poisson bracket equation to ensure $W$ and $P_W$ are canonically conjugate. This is
\begin{equation} \label{28}
\{ W,P_W\} \equiv{\frac{\partial{W}}{\partial{x}}\frac{\partial{P_W}}{\partial{p_x}} - \frac{\partial{W}}{\partial{p_x}}\frac{\partial{P_W}}{\partial{x}}} \equiv{1}
\end{equation}
It is easily solved to yield
\begin{equation} \label{29}
P_W = \frac{1}{dW/dx}p_x = \frac{1}{W}p_x = \frac{dx}{dW}p_x
\end{equation}
or
\begin{equation} \label{30}
P_W = p_x\frac{1}{W} = p_x\frac{dx}{dW}
\end{equation}
(In fact, the $(dW/dx)^{-1}$ can be split between right and left sides of $p_x$. It turns out for Dirac quantization, these are all equivalent under similarity transformations to the above).

Quantizing via Dirac~\cite{Dirac1930-DIRTPO},
\begin{align} \label{31}
\hat{P}_W &= \frac{-i}{(dW/dx)}\frac{d}{dx} = -i(\frac{dx}{dW})\frac{d}{dx} = -i\frac{d}{dW} \\ \label{32}
&= -i\frac{d}{dx}\frac{1}{(dW/dx)} = -i\frac{d}{dx}(\frac{dx}{dW}) = -i\frac{d}{dW}
\end{align}
by the chain rule. This is exactly what we would expect since via Dirac quantization
\begin{equation} \label{33}
\{W,P_W\} = 1 \Rightarrow [\hat{W},\hat{P}_W] = i\hat{1}
\end{equation}
Thus in the $W$-representation,
\begin{equation} \label{34}
\hat{W} = W, \hat{P}_W = -i\frac{d}{dW}
\end{equation}
and in the $P_W$-representation,
\begin{equation} \label{35}
\hat{W} = i\frac{d}{dP_W}, \hat{P}_W = P_W
\end{equation}
It then follows that 
\begin{align} \label{36}
\hat{W}\ket{\Psi_{min}} = -i\hat{K}_W\ket{\Psi_{min}}\\ \label{37}
\hat{W}\ket{\Psi_{0}} = -i\hat{K}_W\ket{\Psi_{0}}
\end{align}
and there is a form-preserving generalized Fourier transform given by
\begin{align} \label{38}
\braket{W}{K_W} = \frac{e^{iWK_W}}{\sqrt{2\pi}}\\ \label{39}
\braket{K_W}{W} = \frac{e^{-iWK_W}}{\sqrt{2\pi}}
\end{align}
(Parenthetically, we note there is another approach that we have also pursued that leads to hidden generalized Harmonic Oscillators \cite{williams2018coupled}! It arises if we write eqs. (\ref{26}) and (\ref{32}) in the $x$-representation:
\begin{equation} \label{40}
\hat{P}_W = \frac{-i}{(dW/dx)}\frac{d}{dx}, \hat{P}_W^{\dagger} = -i\frac{d}{dx}\frac{1}{(dW/dx)}
\end{equation}
These appear not to be self-adjoint (they are termed quasi-Hermitian \cite{scholtz1992quasi}) but in fact, there are. $\hat{P}_W$ must be self-adjoint from eqs. (\ref{31}) and (\ref{32}) and so as an operator, eq. (\ref{40}) must be equivalent but are only manifestly so in the W-representation. Nevertheless, we can still write a Hamiltonian as $\hat{H}$ ($\neq{\hat{H}_W} = \frac{\hat{P}_W^2}{2} + \frac{W^2}{2} $) defined as 
\begin{align} \label{41}
\hat{H} &= \frac{\hat{P}_W^{\dagger}\hat{P}_W}{2} + V(x)\\ \label{42}
&= -\frac{d}{dx}\frac{1}{(dW/dx)^2}\frac{d}{dx} + V(x)
\end{align}
If we try to solve
\begin{equation} \label{43}
\hat{H}\Psi_{0} = E_0\Psi_{0}
\end{equation}
and we set $E_0 \equiv{0}$ then it turns out that
\begin{equation} \label{44}
\hat{H} = \frac{\hat{P}_W^{\dagger}\hat{P}_W}{2} + \frac{1}{2}(W^2 - \frac{dW}{dx}) 
\end{equation}
The other SUSY partner Hamiltonian~\cite{estrada2019ladder} is 
\begin{equation} \label{45}
\hat{H} = \frac{\hat{P}_W\hat{P}_W^{\dagger}}{2} + \frac{1}{2}(W^2 + \frac{dW}{dx}) 
\end{equation}
These lead to the factored forms
\begin{equation} \label{46}
\hat{H}_1 = \frac{1}{\sqrt{2}}(\hat{W} - i\hat{P}_W^{\dagger})\frac{1}{\sqrt{2}}(\hat{W} + i\hat{P}_W)
\end{equation}
and
\begin{equation} \label{47}
\hat{H}_1 = \frac{1}{\sqrt{2}}(\hat{W} + i\hat{P}_W)\frac{1}{\sqrt{2}}(\hat{W} - i\hat{P}_W)
\end{equation}
These result in an entirely new set of generalized Hamiltonians known as Coupled SUSY~\cite{williams2018coupled}. We do not discuss them here because they are valid only if W(x) is a monomial, not a polynomial~\cite{williams2018coupled}.

\section{An Implication of the $\hat{H}_W$ Generalized Harmonic Oscillators for Anomalous and Normal Diffusion}
In several recent studies, we have explored potential applications of the above harmonic oscillators (GHO).

We recently showed that a wide class of anomalous diffusion systems were, in fact, normal diffusion systems when one interpreted the solution to the diffusion equation using the appropriate generalized canonically conjugate position and momentum~\cite{kouri2018point}.

The anomalous diffusion equations of O'Shaughnessy and Procaccia~\cite{o1985analytical} are given by 
\begin{equation} \label{48}
\frac{\partial{\rho}}{\partial{t}} = D\frac{1}{x^{c-1}}\frac{\partial}{\partial{x}}x^{c-1-\vartheta}\frac{\partial{\rho}}{\partial{x}}
\end{equation}
Making the contact transformation
\begin{align} \label{49}
y = W(x) = x^{\beta{C}}\\ \label{50}
\beta = \frac{\vartheta}{2}-c+2\end{align}
the equation becomes 
\begin{equation} \label{51}
\frac{\partial{}\rho}{\partial{t}} = C^2{\beta}^2D\frac{\partial}{\partial{y}}\frac{1}{(\partial{W}/\partial{y})^2}\frac{\partial{\rho}}{\partial{y}},
\end{equation}
where $C$ and $\vartheta$ (or $\beta$ and $C$) are independent of each other. However, there is another new equation
\begin{equation} \label{52}
\frac{\partial{}\rho}{\partial{t}} = C^2{\beta}^2D\frac{1}{(\partial{W}/\partial{y})}\frac{{\partial}^2}{\partial{y}^2}\frac{1}{(\partial{W}/\partial{y})}\rho
\end{equation}
that has not been explored. Eqs. (\ref{51}) and (\ref{52}) can be solved using the appropriate GFT, similar to ordinary normal diffusion. In addition, there results the new diffusion equation
\[\frac{\partial{\rho}}{\partial{t}} = D\frac{{\partial}^2}{\partial{W}^2}\rho\]
with an exact normal diffusion solution:
\begin{equation} \label{53}
\rho{(W(x),t)} = \int \,dW(x')e^{-[W(x) - W(x')]^2/{Dt}}\rho(W(x'),0) 
\end{equation}
(The central limit theorem is clearly valid for the correct dynamical variables)

The anomalous diffusion results from inputting the functions $W(x)$, $W(x)$ explicitly but treating Eq. (\ref{53}) as the time evolution of $\rho{(x,t)}$. That is, the measure is treated as $dx$ rather than $dW(x)$. Calculations for the specific case $W = x + x^3$ resulted in normal $t$ scaling of the MSD at very early times (when $W(x)$ is dominated by $x$) and $t^{1/3}$ at longer times (when $W(x)$ is dominated by $x^3$).

What is this saying? It signifies that depending on the variables considered as canonical, one obtains probability distributions that are pure Gaussian or anomalous! This means that there can be transformations that modify the exponential character of the result! If one applies the standard Fourier transform for $x \rightarrow k $ Eqs. (\ref{51})- (\ref{52}), the result is not Eq. (\ref{53}), a Gaussian solution. Amazingly, there is still a residual central limit result but it is not that of the normal Gaussian.

\section{A Mixed Fourier Transform and Gaussian Invariance}
This suggests we explore more deeply the new GFT kernel
\begin{equation} \label{54}
\braket{W}{K} = \frac{e^{iK(k)W(x)}}{\sqrt{2\pi}}
\end{equation}
If we apply the transform $\braket{W}{K}$ to $e^{-K^2/2}$, the result is $e^{-W^2/2}$. The product $W(x)K(k)$ must be dimensionless. This is easy for monomial $W(x)$ and $K(k)$:
\begin{align} \label{55}
W(x) &= x^{\alpha}cm^{\alpha}\\ \label{56}
K(k) &= k^{\alpha}cm^{\alpha}
\end{align}
and one automatically finds
\begin{equation} \label{57}
\braket{W}{K} = \frac{e^{ix^{\alpha}k^{\alpha}}}{\sqrt{2\pi}}
\end{equation}
Clearly, $WK$ is dimensionless as are $x^{\alpha}$ and $k^{\alpha}$. However, we wish to treat polynomial choices of $W(x)$, $K(k)$, in which vase $W(x)K(k)$ contain cross terms. There are several ways to deal with this. Suppose we take
\begin{equation}
\label{58}
W(x) = \sum_{j=1}^{J}a_jx^{2j-1}
\end{equation}
This is a well-defined Cartesian variable $a_j \geq{0}$, and if $a_1 \neq{0}$, it is invertible for $-\infty < x < \infty$. This is dimensionless if we take $a_j$ to have dimensions 
\begin{equation} \label{59}
a_j \propto{\frac{1}{(cm)^{2j-1}}}
\end{equation}
leading to
\begin{equation}
\label{60}
W(x) = \sum_{j=1}^{J}a_jx^{2j-1},\ a_jx^{2j-1} \ 
\end{equation}
dimensionless.
The same is true for 
\begin{equation}
\label{61}
K(k) = \sum_{j=1}^{J}b_jk^{2j-1},\ b_j\propto{(cm)^{2j-1}}.
\end{equation}
There are other possible choices but for the present, we shall employ this approach. We stress that by this convention, both $W(x)$ and $K(k)$ are each separately dimensionless and each has Cartesian character. (We observe that in the case of even polynomials $W(x)$, $K(k)$,
\begin{align}
W(x) &= \sum_{j=1}^{J}a_jx^{2j},\ a_jx \geq{0}\label{62}
\\ \label{63}
K(k) &= \sum_{j=1}^{J}b_jk^{2j},\ b_j\propto{(cm)^{2j}}
\end{align}
the invertibility condition restricts $x\geq{0}$ and one deals with a radial-like variable.)

For example, let $W(x)$ be the polynomial\cite{williams2018coupled}
\begin{equation}
\label{64}
W(x) = a_1x + a_3x^3
\end{equation}
then
\begin{equation}
\label{65}
\braket{W}{K} = \frac{e^{i(a_1x+a_3x^3)(b_1k+b_3k^3)}}{\sqrt{2\pi}}
\end{equation}
where $(a_1x + a_3x^3)(b_1k + b_3k^3)$ are each dimensionless. However, this means the full transform kernel involves exp$[i(a_1b_1xk + a_3b_1x^k + a_1b_3k^3 + a_3b_3x^3k^3)]$. What do the $xk$, $x^3k^3$, $x^3k^3$ and $xk^3$ portions of the overall transforms do? We easily understand if we explore the action of a general mixed kernel exp$[ix^{\alpha}k^{\beta}]/\sqrt{2\pi}$. Let $y = k^{\beta}$, $z = x^{\alpha}$
\begin{equation} \label{66}
\int_{-\infty}^{\infty} dk^{\beta} \frac{e^{ix^{\alpha}k^{\beta}}}{\sqrt{2\pi}}e^{-k^{2\beta}/2} = \int_{-\infty}^{\infty} dy \frac{e^{izy}}{\sqrt{2\pi}}e^{-y^2/2} = e^{-z^2/2}
\end{equation}
and
\begin{equation} \label{67}
\int_{-\infty}^{\infty} dx^{\alpha} \frac{e^{-ix^{\alpha}k^{\beta}}}{\sqrt{2\pi}}e^{-x^{2\alpha}/2} = \int_{-\infty}^{\infty} dz \frac{e^{-izy}}{\sqrt{2\pi}}e^{-z^2/2} = e^{-y^2/2} = e^{-k^{2\beta/2}}
\end{equation}
Thus, exp$[\pm ix^{\alpha}k^{\beta}]/\sqrt{2\pi}$ transforms between the Gaussian forms: $e^{-k^{2\beta/2}}\leftrightarrow  e^{-x^{2\alpha/2}}$. This implies that each component of the overall kernel provides what is required to convert from a chirp due to $a_1x + a_3x^3$ to a chirp due to $b_1k + b_3k^3$. Individually, they transform from a chirp in $x$ to a harmonic signal in the other variable $K(k)$ and a chirp in $k$. The mixed GFT harmonizes between different types of chirps. This has already been demonstrated for diffusion, where the transform exp$[\pm iWK_W]/\sqrt{2\pi}$ transforms Gaussians in $W$ into Gaussians in $K_W$ and vice-versa. This implies that at the heart of anomalous diffusion is a chirp-based process in the physical displacement $x$. 

\section{Harmonizing Chirp Signals }
We now come to the primary objective of our exploration of mixed Fourier transform kernels. Suppose we are analyzing the signal 
\begin{equation}
\label{68}
S(T) = e^{iT(t)}e^{-T^2(t)/2}
\end{equation}
where
\begin{equation}
\label{69}
T(t) = \sum_{j=1}^{J}a_jt^{2j-1}.
\end{equation}
Thus, it is harmonic in $T$ and a chirp infers of $t$, the physical time. 
	The transform partner $\widetilde{S}(\Omega)$ is given by  
\begin{equation}
\label{70}
\widetilde{S}(\Omega) = \frac{1}{\sqrt{2\pi}}\int_{-\infty}^{\infty} dT \ e^{iT\Omega}S(T)
\end{equation}
where we take
\begin{equation}
\label{71}
\Omega = \sum_{j=1}^{J'}b_jt^{2j-1}.
\end{equation}
We note that $J\neq{J}$, this is a mixed transform from $T(t)$ to $\Omega(\omega)$. The result of the transform is
\begin{align}
\label{72}
\widetilde{S}(\Omega) &= \int_{-\infty}^{\infty} dT \ \frac{e^{iT(\Omega+1)}}{\sqrt{2\pi}}e^{-T^2/2} \\ \label{73}
&= e^{-(\Omega+1)^2/2}.
\end{align}
This is a Gaussian in $\Omega(\omega) = \sum_{j=1}^{J'}b_j\omega^{2j-1}$, centered at $\Omega = -1$. We now carry out another mixed transform from $\Omega$ to the physical time, $t$:
\begin{align}
\label{74}
\hat{S}(t) = \frac{1}{\sqrt{2\pi}}\int_{-\infty}^{\infty}d\Omega \ e^{-i\Omega t}\widetilde{S}(\Omega)\\ \label{75}
\hat{S}(t) = \frac{1}{\sqrt{2\pi}}\int_{-\infty}^{\infty}d\Omega \ e^{-i\Omega t}e^{-(\Omega+1)^2/2}
\end{align}
Let 
\begin{equation}
\label{76}
y = \Omega + 1 \, , dy = d\Omega.
\end{equation}
Then
\begin{align}
\label{77}
\hat{S}(t) &= \frac{1}{\sqrt{2\pi}}\int_{-\infty}^{\infty}dy \ e^{-i(y-1) t}e^{-y^2/2}\\ \label{78}
&= e^{it}\int_{-\infty}^{\infty}dy \ e^{-iyt}e^{-y^2/2}
\\ \label{79}
\hat{S}(t) &= e^{it}e^{-t^2/2}.
\end{align}
This holds for a general choice of $S(T)$ and shows that the use of two appropriate mixed Fourier transforms produces a harmonic signal. The fixed point of interest is the final signal in the omega-domain. This is
\begin{align}
\label{80}
\widetilde{S}(\omega) &= \frac{1}{\sqrt{2\pi}}\int_{-\infty}^{\infty}dt \ e^{-i\omega t}e^{-t^2/2}\\ \label{81}
&= \frac{1}{\sqrt{2\pi}}\int_{-\infty}^{\infty}dt \ e^{it(1-\omega)}e^{-t^2/2}\\ \label{82}
&= e^{-(\omega - 1)^2/2}\end{align}
It yields a Gaussian signal in ordinary frequency centered at $\omega = 1$. This complete the demonstration of harmonizing a chirp using mixed Fourier transforms. 

\section{Conclusions}
This analysis provides a possible procedure to convert non-linear time chirps of the form $\sum_{j=1}^{J}a_jt^{2j-1}$ to the linear signals physical time $t$, and ultimately the physical frequency, $\omega$. It is rigorous and depends solely on generalizing the form of the Fourier transform kernel to one capable of changing a Gaussian in $T$ or $\Omega$ into a Gaussian in $t$ or $\omega$. The original chirp is centered at frequency $\Omega = -1$ and the final harmonic signal is centered at $\omega = \pm 1$. In the case of chirps arising due to anomalous diffusion, we have already shown that this leads to a potential method to determine the appropriate spatial contact transformation $x \rightarrow W(x)$ by careful measurements of the mean square displacement (MSD), $<x^2>$, as a function of time. In the present case of chirps due to transforms $W(x)$, $K(k)$ or $\Omega(\omega)$, $T(t)$, it is not clear what sort of strategy might be used to extract the polynomial transforms experimentally. A brute-force approach would be a search for transforms that harmonize the relevant chirp signal. At present, we do not have a strategy to do this efficiently but it is certainly of significant interest for further research. 
	From a general standpoint, one concludes that chirp signals describable by such point transformations as $W(x)$ or $T(t)$ have a harmonic basis that appears to provide new ideas for the time-frequency or position-wavenumber analysis of chirps. 

%\begin{equation} \label{9}

%\end{equation}

%\label{}
%$ $

% If in two-column mode, this environment will change to single-column format so that long equations can be displayed. 
%\begin{widetext}
%$$\mbox{put long equation here}$$
%\end{widetext}

% Figures should be put into the text as floats. 
% Use the graphics or graphicx packages (distributed with LaTeX2e).
% Put the label that you will use with \ref{} command in the braces of the \label{} command.
%
% \begin{figure}
% \end{figure}

% \begin{table}
% \end{tabular}
% \end{table}

%\begin{acknowledgments}
% Put your acknowledgments here.
%\end{acknowledgments}

\bibliographystyle{plain}
\bibliography{TF_An_Chirps.bib}
\end{document}